\documentclass[12pt]{article}
\usepackage{amsfonts}
\usepackage{amssymb}
\usepackage{graphics,psboxit,amsmath}

\def\hybrid{\topmargin 0pt    \oddsidemargin 0pt
        \headheight 0pt \headsep 0pt
        \textwidth 6.35in       
        \textheight 9.25in       
        \marginparwidth .875in
        \parskip 5pt plus 1pt   \jot = 1.5ex}

\hybrid

\def\baselinestretch{1.2}

\catcode`\@=11

\def\marginnote#1{}
\newcount\hour
\newcount\minute
\newtoks\amorpm
\hour=\time\divide\hour by60
\minute=\time{\multiply\hour by60 \global\advance\minute by-\hour}
\edef\standardtime{{\ifnum\hour<12 \global\amorpm={am}%
        \else\global\amorpm={pm}\advance\hour by-12 \fi
        \ifnum\hour=0 \hour=12 \fi
        \number\hour:\ifnum\minute<10 0\fi\number\minute\the\amorpm}}
\edef\militarytime{\number\hour:\ifnum\minute<10 0\fi\number\minute}

\def\draftlabel#1{{\@bsphack\if@filesw {\let\thepage\relax
   \xdef\@gtempa{\write\@auxout{\string
      \newlabel{#1}{{\@currentlabel}{\thepage}}}}}\@gtempa
   \if@nobreak \ifvmode\nobreak\fi\fi\fi\@esphack}
        \gdef\@eqnlabel{#1}}
\def\@eqnlabel{}
\def\@vacuum{}
\def\draftmarginnote#1{\marginpar{\raggedright\scriptsize\tt#1}}

\def\draft{\oddsidemargin -.5truein
        \def\@oddfoot{\sl preliminary draft \hfil
        \rm\thepage\hfil\sl\today\quad\militarytime}
        \let\@evenfoot\@oddfoot \overfullrule 3pt
        \let\label=\draftlabel
        \let\marginnote=\draftmarginnote
   \def\@eqnnum{(\theequation)\rlap{\kern\marginparsep\tt\@eqnlabel}%
\global\let\@eqnlabel\@vacuum}  }

\def\preprint{\twocolumn\sloppy\flushbottom\parindent 2em
        \leftmargini 2em\leftmarginv .5em\leftmarginvi .5em
        \oddsidemargin -.5in    \evensidemargin -.5in
        \columnsep .4in \footheight 0pt
        \textwidth 10.in        \topmargin  -.4in
        \headheight 12pt \topskip .4in
        \textheight 6.9in \footskip 0pt
        \def\@oddhead{\thepage\hfil\addtocounter{page}{1}\thepage}
        \let\@evenhead\@oddhead \def\@oddfoot{} \def\@evenfoot{} }

\def\numberbysection{\@addtoreset{equation}{section}
        \def\theequation{\thesection.\arabic{equation}}}

\def\underline#1{\relax\ifmmode\@@underline#1\else
        $\@@underline{\hbox{#1}}$\relax\fi}

\def\titlepage{\@restonecolfalse\if@twocolumn\@restonecoltrue\onecolumn
     \else \newpage \fi \thispagestyle{empty}\c@page\z@
        \def\thefootnote{\fnsymbol{footnote}} }

\def\endtitlepage{\if@restonecol\twocolumn \else \newpage \fi
        \def\thefootnote{\arabic{footnote}}
        \setcounter{footnote}{0}}  

\catcode`@=12
\relax

\def\figcap{\section*{Figure Captions\markboth
        {FIGURECAPTIONS}{FIGURECAPTIONS}}\list
        {Figure \arabic{enumi}:\hfill}{\settowidth\labelwidth{Figure
999:}
        \leftmargin\labelwidth
        \advance\leftmargin\labelsep\usecounter{enumi}}}
 \relax
\def\tablecap{\section*{Table Captions\markboth
        {TABLECAPTIONS}{TABLECAPTIONS}}\list
        {Table \arabic{enumi}:\hfill}{\settowidth\labelwidth{Table
999:}
        \leftmargin\labelwidth
        \advance\leftmargin\labelsep\usecounter{enumi}}}
 \relax
\def\reflist{\section*{References\markboth
        {REFLIST}{REFLIST}}\list
        {[\arabic{enumi}]\hfill}{\settowidth\labelwidth{[999]}
        \leftmargin\labelwidth
        \advance\leftmargin\labelsep\usecounter{enumi}}}
 \relax
\makeatletter
\newcounter{pubctr}
\def\publist{\@ifnextchar[{\@publist}{\@@publist}}
\def\@publist[#1]{\list
        {[\arabic{pubctr}]\hfill}{\settowidth\labelwidth{[999]}
        \leftmargin\labelwidth
        \advance\leftmargin\labelsep
        \@nmbrlisttrue\def\@listctr{pubctr}
        \setcounter{pubctr}{#1}\addtocounter{pubctr}{-1}}}
\def\@@publist{\list
        {[\arabic{pubctr}]\hfill}{\settowidth\labelwidth{[999]}
        \leftmargin\labelwidth
        \advance\leftmargin\labelsep
        \@nmbrlisttrue\def\@listctr{pubctr}}}
 \relax
\makeatother
\newskip\humongous \humongous=0pt plus 1000pt minus 1000pt

\newif\ifdtup

\relax

\def\be{\begin{equation}}
\def\ee{\end{equation}}
\def\ba{\begin{eqnarray}}
\def\ea{\end{eqnarray}}

\def\no{\noindent}

\def\IR{\relax{\rm I\kern-.18em R}}
\def\II{\relax{\rm 1\kern-.35em1}}

\renewcommand{\theequation}{\thesection.\arabic{equation}}
\csname @addtoreset\endcsname{equation}{section}

\def\IR{\relax{\rm I\kern-.18em R}}
\def\inv{^{\raise.15ex\hbox{${\scriptscriptstyle -}$}\kern-.05em 1}}

\begin{document}

\begin{titlepage}
\begin{center}


\vskip 1in

{\LARGE Three-point correlation functions from semiclassical circular strings}
\vskip 0.4in

{\bf Rafael Hern\'andez}
 
\vskip 0.1in

Departamento de F\'{\i}sica Te\'orica I\\
Universidad Complutense de Madrid\\
$28040$ Madrid, Spain\\
{\footnotesize{\tt rafael.hernandez@fis.ucm.es}}

\end{center}

\vskip .4in

\centerline{\bf Abstract}
\vskip .1in
\no
The strong-coupling limit of three-point correlation functions of local operators can be analyzed beyond 
the supergravity regime using vertex operators representing spinning string states. When two 
of the vertex operators correspond to heavy string states having large quantum numbers, while 
the third operator corresponds to a light state with fixed charges, the correlator can be computed in the large 
string tension limit by means of a semiclassical approximation. We study the case when the heavy 
string states are circular string solutions with one $AdS_{5}$ spin and three different angular momenta 
along $S^5$, for several choices of the light string state.

\noindent

\vskip .4in
\noindent

\end{titlepage}
\vfill
\eject

\def\baselinestretch{1.2}


\baselineskip 20pt


\section{Introduction} 
  
Complete resolution of a conformal field theory implies determining the whole spectrum of two and three-point 
correlation functions of primary operators. Higher order correlation functions can then be written in terms of 
these two lower ones. In the case of four-dimensional Yang-Mills with ${\cal N}=4$ supersymmetry the spectrum of planar 
anomalous dimensions of single-trace gauge invariant operators has been exhaustively explored, both in the weak and strong-coupling 
regimes, after the uncovering of integrable structures in the AdS/CFT correspondence \cite{Minahan}-\cite{AFS}. 
It is however unclear whether integrability will also illuminate the evaluation of three-point correlation functions. 
In the weak-coupling regime three-point functions can be evaluated perturbatively \cite{perturbative}. 
In the strong-coupling realm a computation at hand within the AdS/CFT correspondence is that of three-point functions 
for chiral operators, which can be evaluated in the supergravity regime \cite{correlationsugra}. But in general the 
calculation of three-point functions requires dealing with primary operators dual to massive string states, 
which is not a tractable problem within our level of understanding of string theory on $AdS_5 \times S^5$. 
However, a case beyond the the supergravity limit and still reachable from the correspondence should be 
that of non-protected operators with large quantum charges, dual to semiclassical spinning string solutions.

The evaluation on the string theory side of the correspondence of correlation functions of single-trace gauge invariant 
operators is performed by inserting closed string vertex operators in the path integral for the string partion function.
Vertex operators scale exponentially with the energy and the quantum charges of the corresponding string state 
and therefore when the charges are as large as the string tension $\sqrt{\lambda}/2\pi$ the string path integral 
can be evaluated through a saddle point approximation. The leading contribution to the corresponding correlation functions 
is thus governed by a semiclassical string configuration. This observation was employed in \cite{twopoint}-\cite{BT} 
to compute two-point correlation functions. The extension to three-point functions has been recently explored in a series 
of appealing papers \cite{Janik}-\cite{RT}, where two of the vertex operators in the correlation function were taken 
to be semiclassical, or heavy, while the remaining light one was chosen as a massless mode, corresponding to a 
protected chiral state \cite{Janik}-\cite{RT}, or as a massive mode, dual to general non-protected states \cite{RT}. 
  
The leading order contribution to the correlator of three string vertex operators is then dominated in the large 
string tension regime by the semiclassical string trajectory coming from the semiclassical operators. 
The quantum numbers of the heavy vertex operators are much larger than those of the light operator, and thus the 
contribution to the saddle point from the light operator can be neglected. 
Therefore, in order to evaluate $\langle V_{H_1}(x_1) V_{H_2}(x_2) V_L(x_3) \rangle$
it suffices to obtain the leading classical string configuration saturating the correlation function of the two heavy 
vertices, $\langle V_{H_1}(x_1) V_{H_2}(x_2) \rangle$, and then evaluate the contribution of the light vertex operator 
$V_L(x_3)$ on this classical solution, 
\be
\langle V_{H_1}(x_1) V_{H_2}(x_2) V_L(x_3) \rangle \ = V_L(x_3)_{\hbox{\tiny{classical}}} \ .
\ee
This observation was employed in reference \cite{RT} to suggest a general method able to cover diverse choices of either 
massless or massive string states for the light vertex operator $V_L(x_3)$. In this note we will closely follow this proposal to 
explore the case where the classical states associated to the heavy vertex operators in the three-point function are 
circular string solutions rotating with one $AdS_5$ spin and three different angular momenta along $S^5$. 
The remaining part of the letter is organized as follows. In section $2$ we will review some relevant 
features of the corresponding spinning string solutions. In section $3$ we will compute the three-point function 
coefficients for several choices of light vertex operators. We conclude in 
section $4$ with some prospects and remarks.


\section{Circular rotating strings}

Semiclassical circular string solutions rotating with several spins and angular momenta in the $AdS_5 \times S^5$ background were 
analyzed in \cite{JRusso}-\cite{NR}. Following notation in there, it will prove useful to parameterize the 
embedding coordinates of the ten-dimensional background in terms of the global $AdS_5$ and $S^5$ angles,
\ba
Y_1 + i Y_2 \! \! \! & = & \! \! \! \sinh \rho \sin \theta e^{i\phi_1} \  , \: \: Y_3 + i Y_4 = \sinh \rho \cos \theta e^{i\phi_2} \  , 
\: \: Y_5 + i Y_0 = \cosh \rho e^{i \, t} \  ,  \\
X_1 + i X_2 \! \! \! & = & \! \! \! \sin \gamma \cos \psi e^{i\varphi_1} \  , \: \: X_3 + i X_4 = \sin \gamma \sin \psi e^{i\varphi_2} \  , 
\: \: X_5 + i X_6 = \cos \gamma e^{i\varphi_3} \  . 
\ea
The $Y_M$ coordinates are related to the Poincar\'e coordinates in $AdS_5$ through 
\be
Y_m = \frac {x_m}{z} \ , \quad Y_4 = \frac {1}{z} \big( -1 + z^2 + x^m x_m \big) \ , \quad 
Y_5 = \frac {1}{2z} \big( 1 + z^2 + x^m x_m \big) \ ,
\ee
where $x^m x_m = -x_0^2 + x_i x_i$, with $m=0,1,2,3$ and $i=1,2,3$. Euclidean rotation allows 
the classical geodesics to reach the boundary, and comes from continuation of the time-like coordinates to
\be
t_e = i t \ , \quad Y_{0e} = i Y_0 \ , \quad x_{0e} = i x_0 \ .
\ee
The two commuting isometries along the $\phi_i$ directions and the three along $\varphi_i$ allow a general ansatz with two 
spins in $AdS_5$ and three angular momenta in $S^5$ \cite{NR},
\ba
Y_1 + i Y_2 \! \! \! & = & \! \! \! b_1 e^{i w_1 \tau + i k_1 \sigma} \  , 
\: \: Y_3 + i Y_4 = b_2 e^{i w_2 \tau + i k_2 \sigma} \  , \: \: Y_5 + i Y_0 = b_0 e^{i \, t} \  ,  \label{ansatz} \\ 
X_1 + i X_2 \! \! \! & = & \! \! \! a_1 e^{i \omega_1 \tau + i m_1 \sigma} \  , 
\: \: X_3 + i X_4 = a_2 e^{i \omega_2 \tau + i m_2 \sigma}  \  , \: \: 
X_5 + i X_6 = a_3 e^{i \nu \tau} \  ,  \label{ansatzS}
\ea
where
\be
t = \kappa \tau \ , \quad w_a^2 = \kappa^2 + k_a^2 \ , \quad b_0^2 - b_1^2 - b_2^2 = 1 \ ,
\ee
with $a=1,2$, and
\be
\omega_i^2 = m_i^2 + \nu^2 \ , 
\quad a_1^2 + a_2^2 + a_3^2 = 1 \ ,
\ee
with $i=1,2,3$ and $m_3=0$, together with the constraints 
\ba
&& E - \kappa \sum_{a=1}^2 \frac {S_a}{w_a} = \sqrt{\lambda} \kappa \ , 
\quad \sum_{i=1}^3 \frac {J_i}{\omega_i} = \sqrt{\lambda} \ , \label{constraints1} \\
&& 2 \kappa E - 2 \sum_{a=1}^2 w_a S_a - \sqrt{\lambda} \kappa^2 = 2 \sum_{i=1}^3 \omega_i J_i - 
\sqrt{\lambda} \nu^2 \ , \label {constraints2} \\
&& \sum_{a=1}^2 k_a S_a + \sum_{i=1}^3 m_i J_i = 0 \ . \label{constraints3}
\ea
Spins along $AdS_5$ are $S_a = \sqrt{\lambda} b_a^2 w_a$, and the angular momenta along $S^5$ are 
$J_i = \sqrt{\lambda} a_i^2 \omega_i$. When all spins and angular momenta are of the same order of magnitude 
and large, we can solve for $\nu$ and $\kappa$ in the above equations as power series expansions \cite{NR},
\be
\nu^2 = \frac {J^2}{\lambda} - \sum_{i=1}^3 m_i^2 \frac {J_i}{J} + \cdots \ , \quad
\kappa^2 = \frac {J^2}{\lambda} + \frac {1}{J} \big( \sum_{i=1}^3 m_i^2 J_i + 2 \sum_{a=1}^2 k_a^2 S_a \big) + \cdots 
\label{kappanu}
\ee
where we have introduced the total angular momentum $J \equiv J_1 + J_2 + J_3$. The energy is then
\be
E = J + S + \frac {\lambda}{2J^2} \big( \sum_{i=1}^3 m_i^2 J_i + \sum_{a=1}^2 k_a S_a \big) + \cdots 
\ee 
  
In what follows we will simply treat the case of a circular string rotating with a single spin $S$ along $AdS_5$, 
and three different momenta $J_i$ along $S^5$. \footnote{The general case of a circular string with two spins 
along $AdS_5$ and three angular momenta along $S^5$ could in principle be also considered. However it 
leads to 
\[
Y_4 + Y_5 = \cosh (\kappa \tau_e) \cosh \rho_0 + \sin (-i w_2 \tau_e + k_2 \sigma) \sinh \rho_0 \cos \theta_0 \ ,
\]
and thus provides hard to evaluate integrals in the three-point vertices below.}Choosing 
\be
b_0 = \cosh \rho_0 \ , \quad b_1 = \sinh \rho_0 \ , \quad b_2 = 0 \ , \label{ansatzAdS}
\ee
the euclidean continuation in Poincar\'e coordinates of this solution becomes
\ba
x_1 \! \! \! & = & \! \! \! \frac {\cos (- i w_1 \tau_e + k_1 \sigma)}{\cosh (\kappa \tau_e)} \tanh \rho_0 \ , \quad 
x_{0e} = \tanh (\kappa \tau_e) \ ,\label {Poincare1} \\ 
x_2 \! \! \! & = & \! \! \! \frac {\sin ( - i w_1 \tau_e + k_1 \sigma)}{\cosh(\kappa \tau_e)} \tanh \rho_0 \ , \quad 
z = \frac {1}{\cosh (\kappa \tau_e) \cosh \rho_0} \ , \label{Poincare2}
\ea 
The choice $b_0 = \cosh \rho_0$ fixes where the string is located in the radial coordinate of $AdS_5$, 
while rotating in the remaining angular directions.
   
Our analysis along this note can be easily truncated to cover two different configurations. 
Setting $\rho_0=0$ corresponds to the case of a circular string with three different angular momenta $J_i$ along $S^5$,
\ba
Y_5 + i Y_0 \! \! & = & \! \! e^{i \kappa \tau} \ , \quad
X_1 + i X_2  = \sin \gamma_0 \cos \psi e^{i\omega_1 \tau + i m_1 \sigma} \ , \nonumber \\
X_3 + i X_4 \! \! & = & \! \! \sin \gamma_0 \sin \psi e^{i\omega_2 \tau + i m_2 \sigma} \ , \quad 
X_5 + i X_6 = \cos \gamma_0 e^{i\nu \tau} \ . 
\ea 
The case of a string with just a single spin $S$ 
and a single momentum $J$ is 
\be
Y_1 + i Y_2 = \sinh \rho_0 e^{i w \tau + i k \sigma} \ , \quad Y_5 + i Y_0 = \cosh \rho_0 e^{i \kappa \tau} \ , \quad 
X_1 + i X_2 = e^{i \omega \tau + i m \sigma} \ . 
\ee
The contribution of this semiclassical solution to a three-point correlator function 
was also considered in reference \cite{Costa}.


\section{Semiclassical three-point functions}

We will now evaluate the leading contribution in the large string tension limit to a three-point correlation function 
with two complex conjugate heavy vertex operators carrying quantum charges of the order of the string tension 
and one light operator with order one charges. 
  
Conformal invariance completely fixes the dependence on the location of the vertex operators 
in a three-point function, up to some coefficient $C_{123}$. The value of these coefficients can be 
obtained from a convenient choice for the positions $x_1$, $x_2$ and $x_3$, namely $|x_1|=|x_2|=1$ 
and $x_3=0$ \cite{BT,RT}. \footnote{This is indeed the case for the semiclassical trajectory (\ref{Poincare1})-(\ref{Poincare2}) 
at the $\tau_e=\pm \infty$ boundaries.}
Then, as the conformal weights of the heavy operators, 
$\Delta_{H_1}=\Delta_{H_2}$, are much larger than that of the light operator, $\Delta_L$,
\be
\langle V_{H_1}(x_1) V_{H_2}(x_2) V_{L}(0) \rangle = \frac {C_{123}} {|x_1-x_2|^{2\Delta_{H_1}}} \ .
\ee
The three-point correlator reduces to the light vertex operator evaluated on the classical solution 
saturating the two-point correlation function of the heavy operators, 
and the value of ${\cal C}_3 \equiv C_{123}/C_{12}$ can then be determined through 
\be
{\cal C}_3 = c_{\Delta}  V_L(0)_{\hbox{\tiny{classical}}} \ ,
\ee
where $c_{\Delta}$ is the normalization constant of the light vertex operator. In what follows we will 
employ the proposal and conventions in reference \cite{RT} in order to evaluate the normalized three-point 
coefficients ${\cal C}_{3}$. The classical states corresponding 
to the heavy vertex operators will be the circular string solutions described in the previous section, and for the 
light vertex operators we will consider several different choices. 

\subsection{Dilaton operator}

We will first analyze the case of a light vertex chosen to be the massless dilaton operator,
\be
V^{\hbox{\tiny{(dilaton)}}} = (Y_+)^{-\Delta_d} \, (X_z)^j 
\big[ z^{-2} (\partial x_m \bar{\partial}x^m + \partial z \bar{\partial} z ) + \partial X_k \bar{\partial} X_k \big] \ ,
\ee
where $Y_+ \equiv Y_4 + Y_5$, $X_z \equiv X_5 + i X_6$ and the derivatives 
are $\partial \equiv \partial_+$ and $\bar{\partial} \equiv \partial_-$. To leading order in the strong-coupling 
regime the scaling dimension is $\Delta_d=4+j$, where we have denoted by 
$j$ the Kaluza-Klein momentum of the dilaton. There is also a fermionic contribution to the dilaton vertex operator, but 
it is subleading in the large string tension expansion, and thus we can safely take into account only the bosonic terms 
in all the vertex operators that we will consider. 
The corresponding gauge invariant operator on the 
gauge theory side is $\hbox{Tr}(F_{\mu \nu}^2 Z^j + \cdots )$. 

The coefficient of the three-point correlator becomes
\be
{\cal C}_{3}^{\hbox{\tiny{(dilaton)}}} = c_{\Delta}^{\hbox{\tiny{(dilaton)}}} 
\int_{-\infty}^{\infty} d \tau_e \int_0^{2\pi} d \sigma \, (Y_+)^{-\Delta_d} \, (X_z)^j 
\big[ z^{-2} (\partial x_m \bar{\partial}x^m + \partial z \bar{\partial} z ) + \partial X_k \bar{\partial} X_k \big] \ ,
\ee
where the normalization constant of the dilaton vertex operator is \cite{Berenstein}
\be
c_{\Delta}^{\hbox{\tiny{(dilaton)}}} = \frac {2^{-j/2-1}}{\pi^2} (j+3) \ .
\ee
The contribution from the $AdS_5$ piece of the circular string ansatz is just $\kappa^2$,
\be
z^{-2} (\partial x_m \bar{\partial}x^m + \partial z \bar{\partial} z ) = \kappa^2 \ ,
\ee
while from (\ref{ansatzS}) the $S^5$ contribution is
\be
\partial X_k \bar{\partial} X_k = - \nu^2 \ ,
\ee
and thus, using that $Y_+ = 1/z$,
\be
{\cal C}_{3}^{\hbox{\tiny{(dilaton)}}}  = 2 \pi c_{\Delta}^{\hbox{\tiny{(dilaton)}}} 
\tilde{a}^2 \int_{-\infty}^{\infty} d \tau_e \frac {e^{j \nu \tau_e}}{(\cosh(\kappa \tau_e))^{4+j}} \ ,
\ee
where we have defined 
\be
\tilde{a}^2 \equiv \frac {(\kappa^2 - \nu^2) (1-a_1^2-a_2^2)^{j/2}} {(1 + b_1^2)^{2+j/2}} \ ,
\ee
with $\kappa^2$ and $\nu^2$ as in equation (\ref{kappanu}). The integral over $\tau_e$ has been evaluated in \cite{RT}, 
and thus our analysis here follows directly from the discussion in there. In the $\nu=0$ limit the string does 
not rotate in the $(56)$-directions, the angular momenta reduce to $J_1 = |m_1| a_1^2$, $J_2 = |m_2| a_2^2$ and $J_3=0$, and 
the three-point vertex is simply
\be
{\cal C}_{3, \nu=0}^{\hbox{\tiny{(dilaton)}}} = 4 \pi^{3/2} c_{\Delta}^{\hbox{\tiny{(dilaton)}}} 
\frac {(1-a_1^2-a_2^2)^{j/2} }{(4+j)(1+b_1^2)^{2+j/2}}
\frac {\Gamma \big( (j+6)/2 \big) }{\Gamma \big( (j+5)/2 \big)} \kappa \ .
\ee
In the case when $\nu \neq 0$ we find
\be
{\cal C}_{3}^{\hbox{\tiny{(dilaton)}}} = 2^{j+5} \pi c_{\Delta}^{\hbox{\tiny{(dilaton)}}}  \tilde{a}^2 
\frac {b_+^{(4)} F_-^{(4)} + b_-^{(4)} F_+^{(4)}}{(4+j)^2 \kappa^2 - j^2 \nu^2} \kappa \ ,
\ee
where we have defined
\be
b_{\pm}^{(\alpha)} \equiv j + \alpha  \pm \frac {j \nu}{\kappa} \quad \hbox{and} \quad 
F_{\pm}^{(\alpha)} \equiv {}_2F_1 \left( j+\alpha, \frac {b_{\pm}^{(\alpha)}}{2}, 1 + \frac {b_{\pm}^{(\alpha)}}{2}, -1 \right) \ ,
\label{bF}
\ee
with $\alpha=4$. In the limit $j=0$ the coupling is just to the lagrangian, and we get
\be
{\cal C}_{3, j=0}^{\hbox{\tiny{(dilaton)}}} = \frac {8 \pi c_{\Delta}^{\hbox{\tiny{(dilaton)}}}}{3(1+b_1^2)^2} 
\frac {\kappa^2-\nu^2}{\kappa} \ .
\label{Cdilaton}
\ee
Let us now concentrate for compactness in the case when the string is moving just along $S^5$, with 
three different angular momenta. Equation (\ref{Cdilaton}) is then \footnote{Comparison with \cite{RT} is immediate 
if we use the condition implied by the Virasoro constraint on the ansatz (\ref{ansatzS}), 
$\kappa^2 = 2 \sum_{i=1}^3 a_i^2 (\omega_i^2 + m_i^2)$. The three-point function can then be written
\[
{\cal C}_{3, j=0}^{\hbox{\tiny{(dilaton)}}} = \frac {8}{3} \pi c_{\Delta}^{\hbox{\tiny{(dilaton)}}} 
\frac {2a_1^2 m_1^2 + 2a_2^2 m_2^2}{\sqrt{2a_1^2 m_1^2 + 2a_2^2 m_2^2 + \nu^2}} \ .
\]
}
\be
{\cal C}_{3, j=0}^{\hbox{\tiny{(dilaton)}}} = \frac {8}{3} \pi c_{\Delta}^{\hbox{\tiny{(dilaton)}}} 
\frac {\sqrt{\lambda} (2m_1^2 J_1 + 2 m_2^2 J_2)}{J \sqrt{J^2 + \frac {\lambda}{J} \left( m_1^2 J_1 + m_2^2 J_2 \right)}} \ ,
\ee
Recalling now that 
\be
E = \sqrt{J^2 + \frac {\lambda}{J} (m_1^2 J_1 + m_2^2 J_2)} \ ,
\ee
our result extends to the case of three different angular momenta the observation in references \cite{Costa} and 
\cite{RT} that the three-point function is proportional to the derivative with respect to $\lambda$ of the 
strong-coupling limit of the anomalous dimension for the corresponding operator,
\be
\lambda \frac {\partial E}{\partial \lambda} = 
\frac {\lambda (m_1^2 J_1 + m_2^2 J_2)}{2 J \sqrt{J^2 + \frac {\lambda}{J} \left( m_1^2 J_1 + m_2^2 J_2 \right)}} \ .
\ee
A similar argument also holds in the more general case of non-vanishing spin along $AdS_5$.  This behavior 
seems to be a general feature in the case of the light dilaton vertex operator, as argued in \cite{Costa} from a 
renormalization group point of view, or in \cite{RT2} by means of a thermodynamical reasoning. 


\subsection{Primary scalar operator}

The dual to the BMN operator $\hbox{Tr}Z^j$ is the superconformal primary scalar operator, and the 
corresponding vertex operator is \cite{Berenstein,Zarembo,RT}
\be
V^{\hbox{\tiny{(primary)}}} = (Y_+)^{-\Delta_p} (X_z)^j 
\big[ z^{-2} (\partial x_m \bar{\partial}x^m - \partial z \bar{\partial} z ) - \partial X_k \bar{\partial} X_k \big] \ ,
\ee
where the scaling dimension is now $\Delta_p=j$. The cases where the classical solution is a BMN geodesic or a folded 
string rotating in $S^5$ have been considered in \cite{Zarembo}, while that of a folded spin rotating in $AdS_5$ has been 
analyzed in \cite{RT}. In this section we will extend the analysis to the case under study in this note, where the semiclassical 
solution is a circular string rotating in $AdS_5$ with a single spin $S$, and with three different angular momenta along $S^5$. 
The ansatz (\ref{ansatz})-(\ref{ansatzS}) leads now to
\be
{\cal C}_3^{\hbox{\tiny{(primary)}}} = 
2 \pi c_{\Delta}^{\hbox{\tiny{(primary)}}} \tilde{b}^2 \int_{-\infty}^{\infty} d \tau_e 
\frac {e^{j \nu \tau_e}}{(\cosh(\kappa \tau_e))^{j}} {\cal I}(\tau_e) \ ,
\ee
where
\be
\tilde{b}^2 \equiv \left( \frac {1-a_1^2-a_2^2}{1 + b_1^2} \right)^{j/2} \quad \hbox{and} \quad
{\cal I}(\tau_e) \equiv \frac {2 \kappa^2}{\cosh^2(\kappa \tau_e)} - 
\frac {2}{J} ( \sum_{i=1}^2 m_i^2 J_i + k \, S) \ .
\ee
In the $\nu = 0$ limit we get
\be
{\cal C}_{3, \nu=0}^{\hbox{\tiny{(primary)}}} = \pi^{3/2} c_{\Delta}^{\hbox{\tiny{(primary)}}} \tilde{b}^2 
\frac {(j-1) \Gamma(j/2)}{\Gamma((j+3)/2)} \kappa \ .
\ee
When $\nu \neq 0$,
\be
{\cal C}_3^{\hbox{\tiny{(primary)}}} = 
2^{j+4} \pi c_{\Delta}^{\hbox{\tiny{(primary)}}} \tilde{b}^2 
\left( \frac {\kappa^2 \big( b_+^{(2)} F_-^{(2)} + b_-^{(2)} F_+^{(2)} \big)}{(j+2)^2\kappa^2-j^2\nu^2} 
- \frac {\big( b_+^{(0)} F_-^{(0)} + b_-^{(0)} F_+^{(0)} \big)}{8j^2} \right)  \kappa \ ,
\label{Cprimary}
\ee
where $b_{\pm}^{(\alpha)}$ and $F_{\pm}^{(\alpha)}$ are as defined in equation (\ref{bF}), with $\alpha=0,2$. 
It is illuminating to consider the limiting case when the classical trajectories from the heavy vertex 
operators approach BMN geodesics, which correspond to point like-strings. If we take $J_1=J_2=S=0$ and $J_3=\sqrt{\lambda} \kappa$, 
relation (\ref{Cprimary}) simplifies to
\be
{\cal C}_3^{\hbox{\tiny{(primary)}}} = 2^{j+3} \pi \frac {j-1}{(j+1)j} c_{\Delta}^{\hbox{\tiny{(primary)}}} \kappa \ .
\ee
Recalling now the normalization constant for the BPS operator \cite{Zarembo},
\be
c_{\Delta}^{\hbox{\tiny{(primary)}}} = \frac {(j+1)\sqrt{j}}{2^{j+3}\pi N} \sqrt{\lambda} \ ,
\ee
the correlator becomes 
\be
{\cal C}_3^{\hbox{\tiny{(primary)}}} = \frac {1}{N} \sqrt{j} J \ ,
\ee
in agreement with the coefficient for the correlator of three chiral primary operators \cite{correlationsugra}.


\subsection{Singlet massive scalar operator}

Let us now consider the case where the light vertex operator is taken to be a singlet massive scalar operator, made out 
of derivatives of the $S^5$ coordinates \cite{RT,RTvertex},
\be
V^{\hbox{\tiny{(singlet)}}} = (Y_+)^{-\Delta_r} \big( (\partial X_k \partial X_k)  (\bar{\partial} X_l \bar{\partial} X_l) \big)^{r/2} \ , 
\quad \hbox{with } \ r=2 \ , 4 \ , \ldots 
\ee
where the scaling dimension is $\Delta_r = 2 \sqrt{(r-1)} \lambda^{1/4}$. 
When $r=2$ the operator corresponds to a massive string state on the first excited level, and the corresponding 
dual gauge theory operator is contained within the Konishi multiplet. Higher values of $r$ label the remaining $(r-1)$-th 
excited levels in the tower of string states. 
  
Using the contribution in (\ref{ansatz}) for the circular string with three different angular momenta 
along $S^5$ we easily get
\be
V^{\hbox{\tiny{(singlet)}}} = \frac {\kappa^{2r}}{(\cosh(\kappa \tau_e) \cosh \rho_0)^{\Delta_r}} \ ,
\ee
Therefore the coefficient in the three-point function is
\be
{\cal C}_{3}^{\hbox{\tiny{(singlet)}}} = \frac {4 \pi^{3/2} c_{\Delta_r}^{\hbox{\tiny{(singlet)}}} }{\Delta_r (1+b_1^2)^{\Delta_r/2}} 
\frac {\Gamma(\Delta_r/2+1)}{\Gamma((\Delta_r+1)/2)} \kappa^{2r-1} \ ,
\ee
as in \cite{RT} when $\rho_0=0$, and the three-point function behaves also as the $(2r-1)$-th power of the 
level number of the light string state in the correlator.

The $AdS_5$ counterpart of this operator produces again an identical result, because 
\be
\big( (\partial Y_K \partial Y_K) (\bar{\partial} Y_L \bar{\partial} Y_L) \big)^{k/2} = \kappa^{2k} \ , 
\quad \hbox{with } \ k=2 \ , 4 \ , \ldots 
\ee
  
The simple structure of the light vertex contribution in both cases happens because 
the singlet scalar operators are made out of chiral components of the stress tensor, and  thus when evaluated 
on {\em any} classical trajectory they imply a constant result \cite{RT}.


\section{Concluding remarks} 

Exhaustive spectroscopy of anomalous dimensions for single-trace gauge invariant operators and energies for the 
corresponding dual strings rotating in the $AdS_5 \times S^5$ background proved essential in order to uncover the integrable structure 
of the AdS/CFT correspondence. In this sense, extending the study of three-point functions to general correlators 
could contribute to clarify whether integrability should also play a role in the evaluation of three-point correlators, 
and thus in the complete resolution of planar ${\cal N}=4$ Yang-Mills. 

In this note we have employed the general proposal in \cite{RT} in order to deal with heavy vertex operators 
corresponding to semiclassical strings rotating in the $AdS_5 \times S^5$ background, and general light vertices. 
The study of additional spinning string solutions contributing to the heavy vertices, as well as different light vertex operators, 
is a natural extension of the present approach to three-point functions at strong-coupling. An additional  
question is the analysis of quadratic fluctuations around the saddle point approximation. 
Understanding this problem, that could hopefully be treated in generality at least in some restricted 
sector of the theory, should also help to clarify the general structure of three-point correlators.


\vspace{10mm}
\centerline{\bf Acknowledgments}

This work is supported by MICINN through a Ram\'on y Cajal contract and grant FPA2008-04906, and by 
BSCH-UCM through grant GR58/08-910770. 


\newpage







\begin{thebibliography}{99}

\renewcommand{\baselinestretch}{.95}
\normalsize

\bibitem{Minahan} J.~A.~Minahan and K.~Zarembo,
{\em The Bethe-ansatz for ${\cal N} = 4$ super Yang-Mills}, 
JHEP {\bf 0303} (2003) 013
{\tt [arXiv:hep-th/0212208]}.

N.~Beisert, C.~Kristjansen and M.~Staudacher,
{\em The dilatation operator of ${\cal N} = 4$ super Yang-Mills theory}, 
Nucl.\ Phys.\  B {\bf 664} (2003) 131, {\tt [arXiv:hep-th/0303060]}.

N.~Beisert and M.~Staudacher,
{\em The ${\cal N}=4$ SYM Integrable Super Spin Chain}, 
Nucl.\ Phys.\  B {\bf 670} (2003) 439, {\tt [arXiv:hep-th/0307042]}.

\bibitem{Bena} I.~Bena, J.~Polchinski and R.~Roiban, 
{\em Hidden symmetries of the $AdS_5 \times S^5$ superstring},
Phys.\ Rev.\  D {\bf 69} (2004) 046002, {\tt [arXiv:hep-th/0305116]}.

V.~A.~Kazakov, A.~Marshakov, J.~A.~Minahan and K.~Zarembo,
{\em Classical / quantum integrability in AdS/CFT},
JHEP {\bf 0405} (2004) 024, {\tt [arXiv:hep-th/0402207]}.

\bibitem{BDS} N.~Beisert, V.~Dippel and M.~Staudacher,
{\em A novel long range spin chain and planar ${\cal N} = 4$ super Yang-Mills},
JHEP {\bf 0407} (2004) 075, {\tt [arXiv:hep-th/0405001]}.

N.~Beisert and M.~Staudacher,
{\em Long-range $PSU(2,2|4)$ Bethe ansaetze for gauge theory and strings},
Nucl.\ Phys.\  B {\bf 727} (2005) 1, {\tt [arXiv:hep-th/0504190]}.

\bibitem{AFS} G.~Arutyunov, S.~Frolov and M.~Staudacher, 
{\em Bethe ansatz for quantum strings},
JHEP {\bf 0410} (2004) 016, {\tt [arXiv:hep-th/0406256]}.

N.~Beisert, 
{\em The $su(2|2)$ dynamic S-matrix},
Adv.\ Theor.\ Math.\ Phys.\  {\bf 12} (2008) 945, {\tt [arXiv:hep-th/0511082]}.

R.~A.~Janik,
{\em The $AdS_5 \times S^5$ superstring worldsheet S-matrix and crossing  symmetry}, 
Phys.\ Rev.\  D {\bf 73} (2006) 086006, {\tt [arXiv:hep-th/0603038]}.

N.~Beisert, R.~Hern\'andez and E.~L\'opez,
{\em A crossing-symmetric phase for $AdS_5 \times S^5$ strings},
JHEP {\bf 0611} (2006) 070, {\tt [arXiv:hep-th/0609044]}.

N.~Beisert, B.~Eden and M.~Staudacher,
{\em Transcendentality and crossing}, 
J.\ Stat.\ Mech.\  {\bf 0701} (2007) P021, 
{\tt [arXiv:hep-th/0610251]}.

\bibitem{perturbative} C.~Kristjansen, J.~Plefka, G.~W.~Semenoff and M.~Staudacher,
{\em A new double-scaling limit of ${\cal N} = 4$ super Yang-Mills theory and PP-wave strings},
Nucl.\ Phys.\  B {\bf 643} (2002) 3, {\tt [arXiv:hep-th/0205033]}.

N.~R.~Constable, D.~Z.~Freedman, M.~Headrick, S.~Minwalla, L.~Motl, A.~Postnikov and W.~Skiba,
{\em PP-wave string interactions from perturbative Yang-Mills theory},
JHEP {\bf 0207} (2002) 017, {\tt [arXiv:hep-th/0205089]}.

C.~S.~Chu, V.~V.~Khoze and G.~Travaglini,
{\em Three-point functions in ${\cal N} = 4$ Yang-Mills theory and pp-waves},
JHEP {\bf 0206} (2002) 011, {\tt [arXiv:hep-th/0206005]}.

N.~Beisert, C.~Kristjansen, J.~Plefka, G.~W.~Semenoff and M.~Staudacher,
{\em BMN correlators and operator mixing in ${\cal N} = 4$ super Yang-Mills theory},
Nucl.\ Phys.\  B {\bf 650} (2003) 125, {\tt [arXiv:hep-th/0208178]}.

K.~Okuyama and L.~S.~Tseng,
{\em Three-point functions in ${\cal N} = 4$ SYM theory at one-loop},
JHEP {\bf 0408} (2004) 055, {\tt [arXiv:hep-th/0404190]}.

L.~F.~Alday, J.~R.~David, E.~Gava and K.~S.~Narain,
{\em Structure constants of planar ${\cal N} = 4$ Yang Mills at one-loop},
JHEP {\bf 0509} (2005) 070, {\tt [arXiv:hep-th/0502186]}.

G.~Georgiou, V.~L.~Gili and R.~Russo,
{\em Operator mixing and three-point functions in ${\cal N}=4$ SYM},
JHEP {\bf 0910} (2009) 009, {\tt arXiv:0907.1567 [hep-th]}.

A.~Grossardt and J.~Plefka,
{\em One-Loop Spectroscopy of Scalar Three-Point Functions in planar ${\cal N}=4$ super Yang-Mills Theory},
{\tt arXiv:1007.2356 [hep-th]}.

\bibitem{correlationsugra} D.~Z.~Freedman, S.~D.~Mathur, A.~Matusis and L.~Rastelli,
{\em Correlation functions in the $CFT_d/AdS_{d+1}$ correspondence},
Nucl.\ Phys.\  B {\bf 546} (1999) 96, {\tt [arXiv:hep-th/9804058]}.

G.~Chalmers, H.~Nastase, K.~Schalm and R.~Siebelink,
{\em R-current correlators in N = 4 super Yang-Mills theory from anti-de  Sitter supergravity},
Nucl.\ Phys.\  B {\bf 540} (1999) 247, {\tt [arXiv:hep-th/9805105]}.

S.~Lee, S.~Minwalla, M.~Rangamani and N.~Seiberg,
{\em Three-point functions of chiral operators in $D = 4$, ${\cal N} = 4$ SYM at  large $N$},
Adv.\ Theor.\ Math.\ Phys.\  {\bf 2} (1998) 697, {\tt [arXiv:hep-th/9806074]}.

G.~Arutyunov and S.~Frolov,
{\em Some cubic couplings in type IIB supergravity on $AdS_5 \times S^5$ and three-point functions in $SYM(4)$ at large $N$},
Phys.\ Rev.\  D {\bf 61} (2000) 064009, {\tt [arXiv:hep-th/9907085]}.

S.~Lee,
{\em $AdS_5/CFT_4$ Four-point Functions of Chiral Primary Operators: Cubic Vertices},
Nucl.\ Phys.\  B {\bf 563} (1999) 349, {\tt [arXiv:hep-th/9907108]}.

\bibitem{twopoint} S.~S.~Gubser, I.~R.~Klebanov and A.~M.~Polyakov,
{\em A semi-classical limit of the gauge/string correspondence},
Nucl.\ Phys.\  B {\bf 636} (2002) 99, {\tt [arXiv:hep-th/0204051]}.

\bibitem{twopointTseytlin} A.~A.~Tseytlin,
{\em On semiclassical approximation and spinning string vertex operators in $AdS_5 \times S^5$},
Nucl.\ Phys.\  B {\bf 664} (2003) 247, {\tt [arXiv:hep-th/0304139]}.

\bibitem{Buchbinder} E.~I.~Buchbinder,
{\em Energy-Spin Trajectories in $AdS_5 \times S^5$ from Semiclassical Vertex Operators},
JHEP {\bf 1004} (2010) 107, {\tt arXiv:1002.1716 [hep-th]}.

\bibitem{BT} E.~I.~Buchbinder and A.~A.~Tseytlin,
{\em On semiclassical approximation for correlators of closed string vertexoperators in AdS/CFT}, 
JHEP {\bf 1008} (2010) 057, {\tt arXiv:1005.4516 [hep-th]}.

\bibitem{Janik} R.~A.~Janik, P.~Surowka and A.~Wereszczynski,
{\em On correlation functions of operators dual to classical spinning string states},
JHEP {\bf 1005} (2010) 030, {\tt arXiv:1002.4613 [hep-th]}.

\bibitem{Costa} M.~S.~Costa, R.~Monteiro, J.~E.~Santos and D.~Zoakos,
{\em On three-point correlation functions in the gauge/gravity duality},
{\tt arXiv:1008.1070 [hep-th]}.

\bibitem{Zarembo} K.~Zarembo,
{\em Holographic three-point functions of semiclassical states},
JHEP {\bf 1009} (2010) 030, {\tt arXiv:1008.1059 [hep-th]}. 

\bibitem{RT} R.~Roiban and A.~A.~Tseytlin,
{\em On semiclassical computation of 3-point functions of closed string vertex operators in $AdS_5 \times S^5$}, 
{\tt arXiv:1008.4921 [hep-th]}.

\bibitem{JRusso} J.~G.~Russo, 
{\em Anomalous dimensions in gauge theories from rotating strings in $AdS_5 \times S_5$},
JHEP {\bf 0206} (2002) 038, {\tt [arXiv:hep-th/0205244]}.

\bibitem{FT} S.~Frolov and A.~A.~Tseytlin, 
{\em Multi-spin string solutions in $AdS_5 \times S^5$},
Nucl.\ Phys.\  B {\bf 668} (2003) 77, {\tt [arXiv:hep-th/0304255]}.

\bibitem{AFRT} G.~Arutyunov, S.~Frolov, J.~Russo and A.~A.~Tseytlin,
{\em Spinning strings in $AdS_5 \times S^5$ and integrable systems},
Nucl.\ Phys.\  B {\bf 671} (2003) 3, {\tt [arXiv:hep-th/0307191]}.

\bibitem{NR} G.~Arutyunov, J.~Russo and A.~A.~Tseytlin,
{\em Spinning strings in $AdS_5 \times S^5$: New integrable system relations},
Phys.\ Rev.\  D {\bf 69} (2004) 086009, {\tt [arXiv:hep-th/0311004]}.

\bibitem{Berenstein} D.~E.~Berenstein, R.~Corrado, W.~Fischler and J.~M.~Maldacena,
{\em The operator product expansion for Wilson loops and surfaces in the  large $N$ limit},
Phys.\ Rev.\  D {\bf 59} (1999) 105023, {\tt [arXiv:hep-th/9809188]}.

\bibitem{RT2} R.~Roiban and A.~A.~Tseytlin,
{\em Spinning superstrings at two loops: strong-coupling corrections to dimensions of large-twist SYM operators},
Phys.\ Rev.\  D {\bf 77} (2008) 066006, {\tt arXiv:0712.2479 [hep-th]}.

\bibitem{RTvertex} R.~Roiban and A.~A.~Tseytlin,
{\em Quantum strings in $AdS_5 \times S^5$: strong-coupling corrections to dimension of Konishi operator},
JHEP {\bf 0911} (2009) 013, {\tt arXiv:0906.4294 [hep-th]}.


\end{thebibliography}
\end{document}